\documentclass[useAMS,usenatbib]{mn2e}
\usepackage{natbib}
\usepackage{amsmath}
\usepackage{subfigure}
\usepackage{gensymb}
\usepackage{amssymb}
\usepackage{wasysym}
\usepackage{graphicx}
\usepackage{color}
\usepackage{tabularx}
\usepackage{xcolor}
\usepackage{ulem}

\def\memb{M_{\rm emb}}

\begin{document}

\title[Limits on protoplanets in the primordial disk]{Limits on the number of primordial Scattered Disk objects at Pluto mass and higher from the absence of their dynamical signatures on the present day trans-Neptunian Populations}
\author[Shannon \& Dawson]{Andrew Shannon$^{1,2}$ \& Rebekah Dawson$^{1,2}$ \\
$^{1}$Department of Astronomy \& Astrophysics, The Pennsylvania State University, State College, PA, USA \\
$^{2}$Center for Exoplanets and Habitable Worlds, The Pennsylvania State University, State College, PA, USA}

\maketitle

\begin{abstract}
Today, Pluto and Eris are the largest and most massive Trans-Neptunian Objects respectively.  They are believed to be the last remnants of a population of planetesimals that has been reduced by $>99\%$~since the time of its formation.  This reduction implies a primordial population of hundreds or thousands of Pluto-mass objects, and a mass-number distribution that could have extended to hundreds of Lunas, dozens of Mars, and several Earths.  Such lost protoplanets would have left signatures in the dynamics of the present-day Trans-Neptunian Populations, and we statistically limit their primordial number by considering the survival of ultra-wide binary TNOs, the Cold Classical Kuiper belt, and the resonant population. 
We find that if the primordial mass-number distribution extended to masses greater than Pluto $\left(\sim 10^{-3} M_\oplus\right)$,
it must have turned downwards to be no more top-heavy than roughly equal mass per log size, a significant deviation from the distribution observed between $10^{-5} M_\oplus$~and $10^{-3}~M_\oplus$. We compare these limits to the predicted mass-number distribution of various planetesimal and proto-planet growth models. The limits derived here provide a test for future models of planetesimal formation.
\end{abstract}


\section{Introduction}

Beyond Neptune lie remnant small bodies from the planet formation era that were never incorporated into planets.  While some of these Trans-Neptunian objects, known as Cold Classical Kuiper belt objects, are thought to have formed more-or-less in situ (e.g., \citealt{park12,baty12,daws12}), the other bodies (which we will call Hot Trans-Neptunian objects [Hot TNOs], although numerous naming schemes and sub-groups are  used\footnote{the Hot population is sometimes divided into sub-populations such as Hot Classicals, Scattering, Detached, Extended-Detached, but they are generally believed to arise from the same source population \citep[e.g.,][]{levi08,2017AJ....153..127P}. 
Hot Classicals are sometimes bundled with Cold Classicals as the Classical population.}) are thought to have been moved to their current orbits from initial orbits which were closer to the sun by the actions of the planets \citep{malh93,malh95,chia02,levi03,levi08,nesv16a}.  

While today the largest TNOs are Pluto and Eris, with radii $\sim 1000 \rm{km}$, some models of planetesmal formation and growth predict that Earth and Mars mass objects should have formed at 20--30 au (e.g., \citealt{joha15,hopk16}) where the Hot TNOs are believed to have formed.  Other models predict planetesimals should only grow to Pluto mass (e.g., \citealt{keny98,orme10,schl11,shan15}), and for other models, the input conditions determine the maximum mass (e.g.,  \citealt{keny12,cham16}).  Thus, reconstruction of the primordial size-number distribution can be a useful tool for discriminating among planetesimal formation and growth models.

Several arguments suggest hundreds to thousands of Pluto-mass objects (hereafter, Plutos) may have existed in the primordial outer Solar system \citep{ster91}.  The creation of the Pluto-Charon binary in a giant impact \citep{mcki89} requires a large number of Plutos to have a reasonable chance of occurring.  The capture of Triton by Neptune \citep{agno06,nogu11} also requires a substantial number of Plutos to have a reasonable chance of occurring. A large number of Plutos can cause stocasticity in Neptune's outward migration and keep the resonance capture fraction low \citet{murr06}. \citet{murr06} argued that the present day resonant fraction implies an upper limit that at most a few percent of the primordial mass was in Plutos, whereas \citet{nesv16} argued that the present day resonant fraction is low and requires stochasticity introduced by 1000-4000 Plutos.

It is unclear whether a substantial number of objects larger than Pluto were present.  If the obliquities of Uranus and Neptune are the result of their last big impact, those impactors would have been Earths \citep{pari11,pari11a}, but this idea is speculative, and alternative origins for their obliquities have been proposed \citep{kubo95,boue10,morb12}.  \citet{2015A&A...582A..99I} demonstrated that Uranus and Neptune could be produced from $\sim 2--6$ Earth mass protoplanets.  The existence of the detached disk, Hot TNOs with pericentre $q >> 40 \rm{au}$~, can be explained if Mars-to-Earth mass objects were in the trans-Neptunian region and raised Hot TNOs, resulting in higher pericentres \citep{glad06}.  Tens of such bodies would transfer about $1\%$~of the Hot TNOs to the detached disk \citep{sils17}.  The current estimate of the total mass in detached objects is $\sim 0.1 M_{\oplus}$~\citep{shep16}. It is difficult to estimate the uncertainty in that total mass because of uncertainty in the eccentricity distribution of the extended disk \citep[which depends on the Solar system's history, ][]{lawl17}, and the small number of detections poorly constraining the size distribution slope and normalisation.

The primordial mass-number distribution of the planetesimals and protoplanets will impact the evolution of the Solar system. Many investigations have focused on the effects of giant planets, through which different Solar system histories of giant planets will leave different imprints in the dynamics of the Hot TNOs \citep[e.g.,][]{gome03,murr05,murr11,nesv15,park15}, as well as on the Cold Classicals \citep{baty11,murr11,daws12,nesv15}.  Reversing this thinking, some authors have used the observed features of the Trans-Neptunian population to infer the existence of now-lost planets \citep{glad06,lyka08,sils17}, or ever as-yet-undiscovered planets \citep{baty16}. Just as we might hypothesise the existence now-lost planet(s) to explain observed features of the Trans-Neptunian populations, we can hypothesise the non-existence of now-lost planets to explain the lack of observed features in the Trans-Neptunian populations.  For instance, \citet{morb02} considered the retention fraction of Lunas, Mars, and Earths that formed in the Cold Classical region, and argued that had more than a couple formed, we would still expect to see at least one today.  In that vein, we discuss how the present day population evolves from the primordial population of TNOs originally formed at 20--30 au in \textsection \ref{sec:pastnpresent}, and then consider three observed properties of the Trans-Neptunian populations to constrain the primordial populaton:

1) A large fraction of the ultra-wide binaries in the Cold Classical Kuiper belt survived to the the present day (\textsection \ref{sec:BinarySplits}).

2) The Cold Classical Kuiper belt has remained dynamically unexcited (\textsection \ref{sec:coldexcite}).

3) A fraction of the resonant objects must survive until the present day (\textsection \ref{sec:reslives}).  

We discuss the implication of these limits for models of planetesimal and protoplanet formation and growth, as well as for the capture of resonant objects and the origin of Triton, in \textsection \ref{sec:discussion}.  We summarise our conclusions in \textsection \ref{sec:conclusions}.

\section{Observations and our inferences of the Past}
\label{sec:pastnpresent}

\begin{figure*}
\includegraphics[width=0.27\textwidth, angle=270, trim = 0 50 0 100, clip]{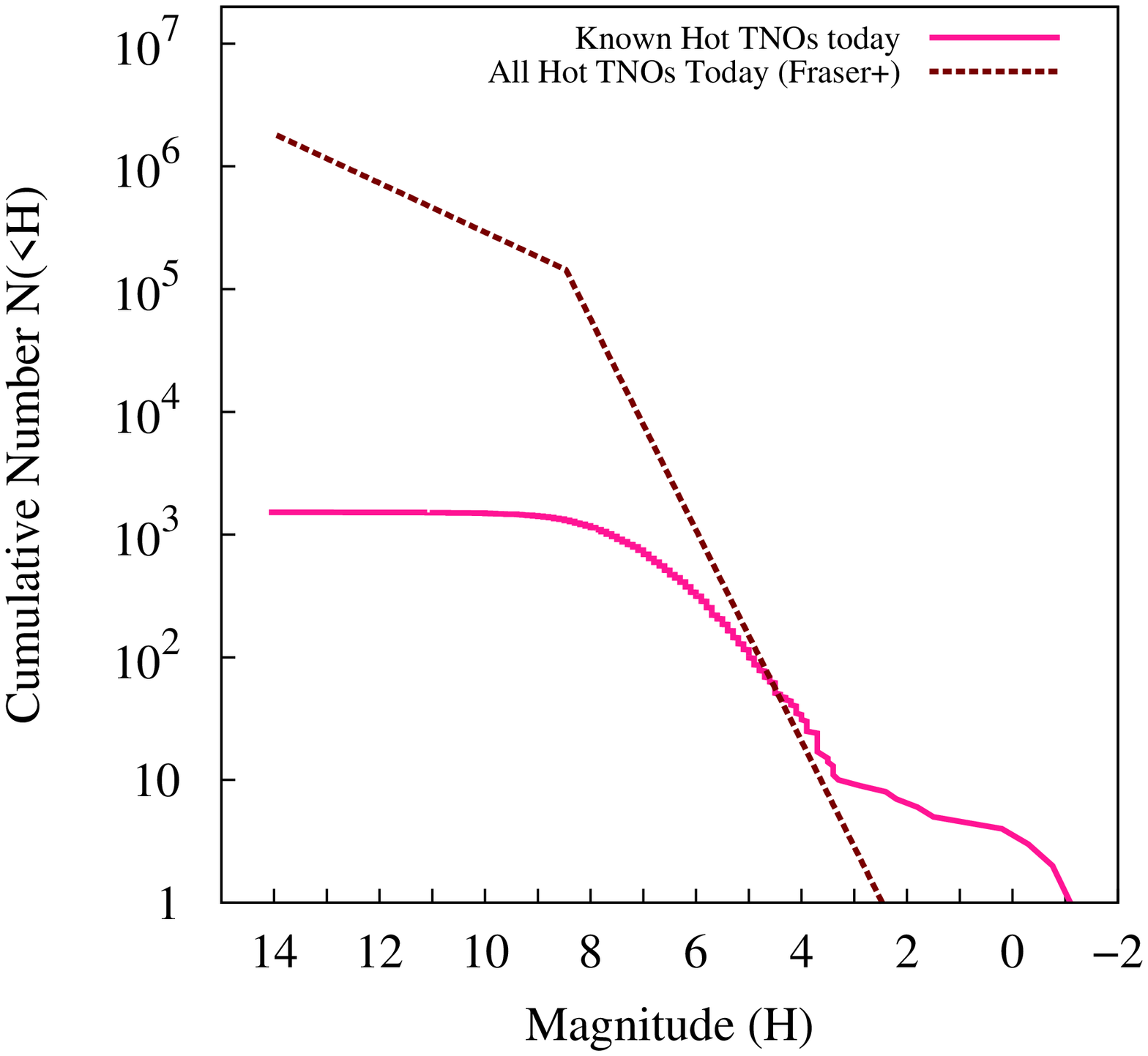}
\includegraphics[width=0.27\textwidth, angle=270, trim = 0 50 0 100, clip]{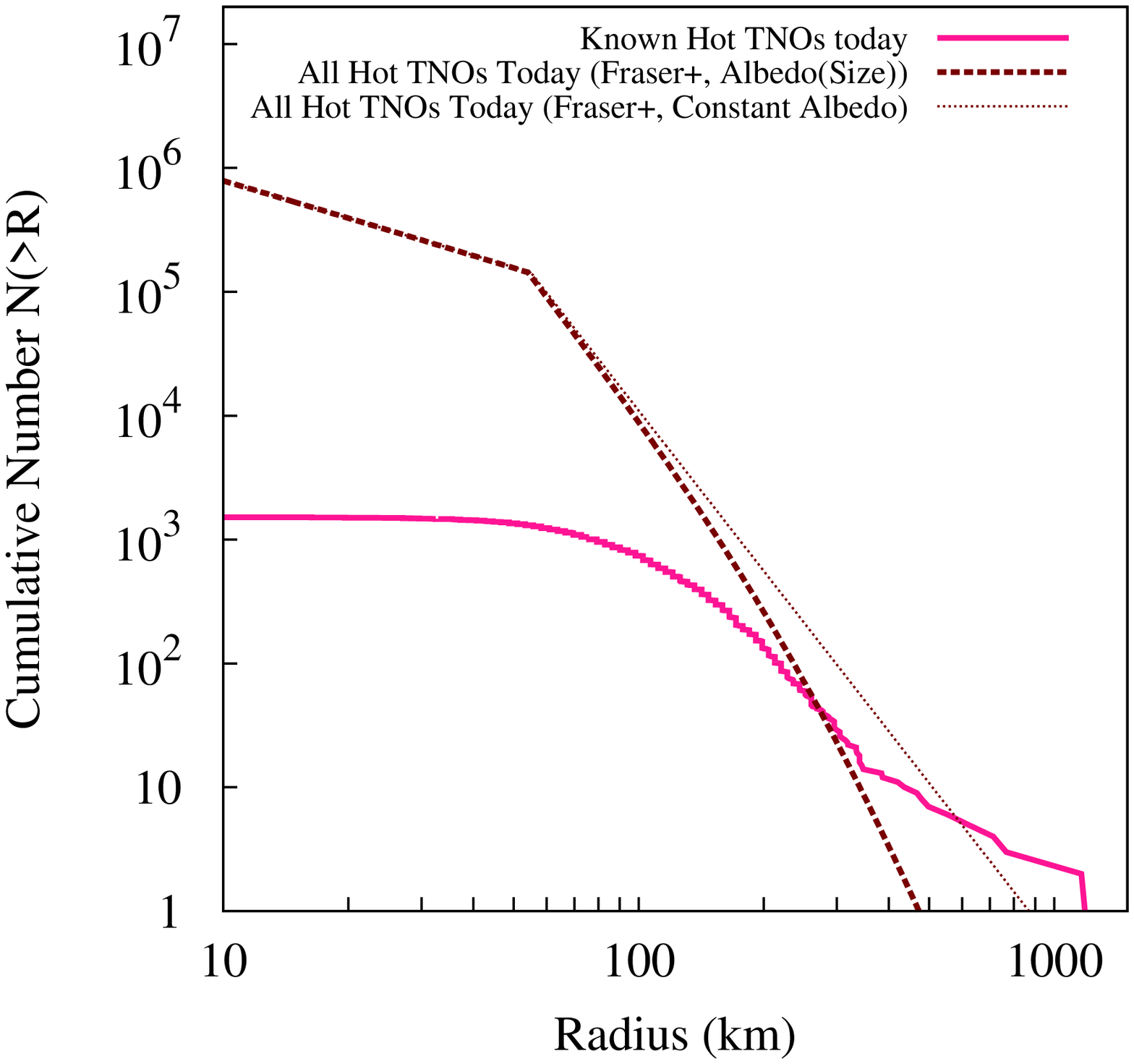}
\includegraphics[width=0.27\textwidth, angle=270, trim = 0 50 0 00, clip]{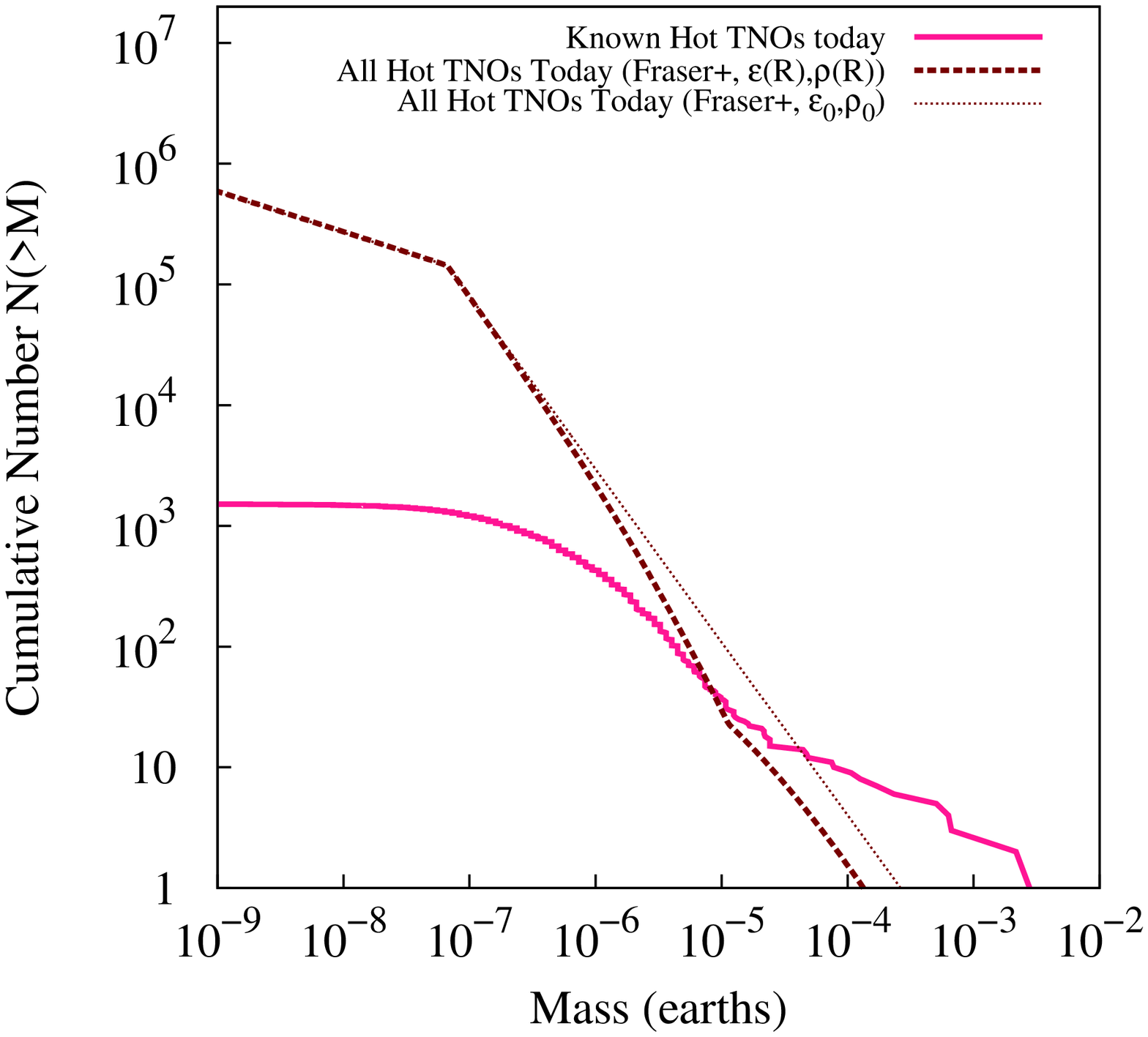}
  \caption{Known Hot Trans-Neptunian Objects, compared with the completeness corrected population of \citet{2014ApJ...782..100F}.  The brightest object in the \citet{2014ApJ...782..100F} sample has absolute magnitude $H \sim 5$, at brighter magnitudes the relation is simply the extrapolation.  For sizes, the known objects use radii measured by occultations, direct imaging, thermal models, or assumed albedo, in that order of preference (sources in text).  The thick \citet{2014ApJ...782..100F}~line is the size distribution with the albedo-size relation found in that work, while the thin one is for a constant albedo.  For masses, we used measured masses from binaries where available (sources in text) where available, otherwise an assumed density of $\rho = 0.6 \rm{g/cm^3}$~for objects with $R < \rm{300 km}$, $\rho = 2.0 \rm{g/cm^3}$~for objects with $R > 500~\rm{km}$, and densities linearly interpolated at intermediate sizes.  The \citet{2014ApJ...782..100F}~masses use the same assumed densities, and the \citet{2014ApJ...782..100F}~size-albedo (thick line), and a constant albedo and density (think line).   }
  \label{fig:empirical}
\end{figure*}

To understand the implications of the limits we will derive, we can compare them both to predictions from planet formation theory and to limits to the primordial population implied by observed TNOs.  Both of these require a general understanding of how the primordial population evolves to the present.  In this section we discuss relevant properties of the present day observed population, how we connect the primordial population to the present day, and what the present day population implies about the primordial population.

In figure \ref{fig:empirical} we plot size distributions of hot TNOs observed today. We plot the size distribution inferred by \citet{2014ApJ...782..100F} from a joint re-analyses of the surveys of \citet{1998AJ....116.2042G,2001ApJ...549L.241A,2001AJ....122..457T,2004AJ....128.1364B,2008AJ....136...83F,peti11}, which allowed for well determined distances, and the completeness to be accurately characterised down to radii of $\sim 15 \rm{km}$.  The survey found no objects with $R \gtrsim 250 \rm{km}$ in the search area.  There have been tentative reports of a change to a shallower size-number distribution among the largest bodies in the hot TNO population \citep{adam14,2014ApJ...782..100F}.  Considering this, we also plot all known objects.  The observed population is taken from the Minor Planet Center (on 10th October 2017), with measured radii from occulations \citep{elli10,sica11,brag13,alva14,dias17,schi17}, direct imaging \citep{nimm17}, thermal modelling \citep{lim10,mull10,momm12,pal12,sant12,vile12,2013A&A...555A..15F,lell13,vile14}, or by assuming albedos of $\left[\left(D/\rm{250~km}\right)^2+6\right]\%$~per \citet{2014ApJ...782..100F}, in that order of priority.  Masses for objects with satellites were taken from \citet{2010Icar..207..978B,2013ApJ...778L..34B,2007Sci...316.1585B,2011A&A...534A.115C,2013Icar..222..357F,2007Icar..191..286G,2008Icar..197..260G,2009Icar..200..627G,2011Icar..213..678G,2012Icar..220...74G,2015Icar..257..130G,2017A&A...608A..19K,2004DPS....36.0803M,park11,2009AJ....137.4766R,2012AJ....143...58S,2012Icar..219..676S,2015Sci...350.1815S,2002Natur.416..711V}.  To estimate masses for other TNOs, we assign a density following the data in \citet{2013ApJ...778L..34B}, with $\rho = 0.6~\rm{g/cm^{3}}$~for objects with $R < 300~\rm{km}$, $\rho = 2.0~\rm{g/cm^{3}}$~for objects with $R > 500~\rm{km}$, and connect them linearly in between, i.e., $\rho = 1.4\times(R-300~\rm{km})/200~\rm{km} + 0.6~\rm{g/cm^{3}}$.  This formula would give a density of $0.91~\rm{g/cm^3}$~to 2002 UX$_{25}$, with measured density $0.82 \pm 0.11~\rm{g/cm^3}$~\citep{2013ApJ...778L..34B}, a density of $1.86~\rm{g/cm^3}$~to Orcus with measured density $1.53^{+0.15}_{-0.13}~\rm{g/cm^3}$~\citep{2013A&A...555A..15F}, and a density of $1.55~\rm{g/cm^3}$~to Salacia, with measured density $1.29^{+0.29}_{-0.23}~\rm{g/cm^3}$~\citep{2013A&A...555A..15F}, the objects with known density in that size range.  There is an apparent change to a shallower magnitude-number, size-number, and mass-number distribution among the known objects at a radius of $\sim 400$ km (figure \ref{fig:empirical}).  That change
contains a host of poorly known sensitivities, and as such should be viewed with scepticism, but the change in slope is the opposite of the naive expectations for detection biases, which typically favour the discovery of larger objects.   Since observed size distribution matches the \citet{2014ApJ...782..100F} distribution at $R \sim 250 \rm{km}$~the known sample at $250+ \rm{km}$~can be reasonably supposed to be complete within a factor of a few. Indeed, all sky surveys for very large KBOs \citep[e.g.,][]{2014AJ....147....2S,2015AJ....149...69B} find that the known objects are mostly complete above $R \sim 500 \rm{km}$. 
Thus, it would be difficult to make more than a minor modification to the large-size end of the size-number distribution based on observational completeness or to explain the apparent change in the slope of the size-number distribution at $R \sim 250 \rm{km}$~by completeness arguments. Completeness should produce the opposite change from that observed (i.e., if the entire distribution truly had the same negative slope, completeness would make the apparent slope shallower at smaller sizes).  The present day mass-number distribution of objects is $N\left(>M\right) \sim 1$~near Pluto's mass, which naturally suggests the possibility it may have extended to significantly larger mass in the past, when the normalisation was orders of magnitude larger.

To scale the present day population to an inferred initial population, we perform $N$-body simulations of the evolution of the outer Solar system.  All $N$-body simulations performed in this study use {\tt mercury6} \citep{cham99} with the hybrid symplectic integrator and a timestep of 120 days. 

In our nominal scenario, the giant planets expand from a compact configuration, scattering out a primordial planetesimal disk \citep{fern84,malh93,malh95} -- which extended to roughly 30 au \citep{gome04} -- that becomes today's Hot Trans-Neptunian population \citep{levi08}. We begin with Jupiter at 5.8 au, Saturn at 8.6 au, Uranus at 12.9 au, Neptune at 18 au, and 1000 planetesimals distributed from 18-30 au according to surface density profile $\Sigma \propto a^{-1}$, all on circular orbits. Our simulated planetesimals have individual masses of $2\times10^{26}$g (0.3 Mars masses; a total mass of 33.3 $M_{\oplus}$) and interact gravitationally with the planets but not each other. We repeat the simulation three times to assess whether the results are sensitive to stochastic differences in the evolution. In all three cases, Neptune, Uranus, and Saturn migrate out and Jupiter migrates in to roughly their current orbits.  

Over the course of the 4 Gyrs evolution,
about half ($51.7\%$, $55.4\%$, and $44.7\%$ in the three trials) of primordial disk objects formed interior to 30 AU are ever members of the Hot Trans-Neptunian population, defined here as $q > 15 \rm{au}$~and $a > 35 \rm{au}$. (The other half of objects are ejected onto hyperbolic orbits, ejected at Jupiter or Saturn crossing orbits $q \sim 5-10$, or hit a planet.)

\begin{figure}
\includegraphics[width=0.33\textwidth, angle=270, trim = 0 0 0 0, clip]{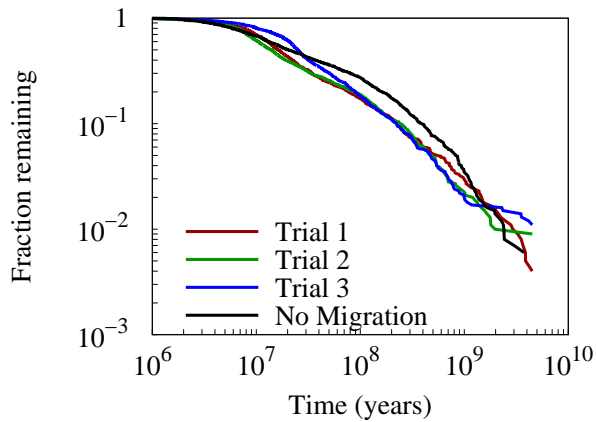}
  \caption{Fraction of objects scattered into the Kuiper belt that remain in Kuiper belt from three simulations (red, green, and blue lines) of our nominal dynamical history scenario, where the giant planets start more compact, and expand to their current orbits, and a comparison simulation where they begin on their present orbits and do not migrate (black line).  The lifetime distributions of trans-Neptunian objects are similar in all cases.  
  }
  \label{fig:scatlifetime}
\end{figure}

To assess whether intrusion of Hot TNOs into the Cold Classical region is sensitive to the assumed dynamical history of the Solar System, we compare our nominal scenario to another scenario. 
In the second scenario, the giant planets are situated on their current orbits, and the simulated planetesimals begin on the same orbits as the migration case, but with zero mass.  The simulations produce very similar results, for instance we plot the lifetimes of Hot TNOS from our nominal solar system dynamical history (figure \ref{fig:scatlifetime}).  The result is in good agreement with other investigations, such as \citet{dunc97} who first investigated the lifetimes of objects scattered in the Kuiper belt and found that $\sim 1\%$~of objects persisted for $4{\rm Gyr}$. 

These simulations of dynamical histories will be critical to our constraints on the primordial population, because the trans-Neptunian populations were exposed to a population of scattered bodies that decreased over time.  We can invert the survival fraction of scattered bodies to transform the present day population into an estimate of the primordial population.  Additionally, we can check that the limits we derive on the number of primordial Pluto-mass bodies are compatible with the number of Plutos observed today scaled by the inverse survival fraction (\textsection \ref{sec:discussion}). 

\begin{figure}
\includegraphics[width=0.42\textwidth, angle=270, trim = 0 75 0 100, clip]{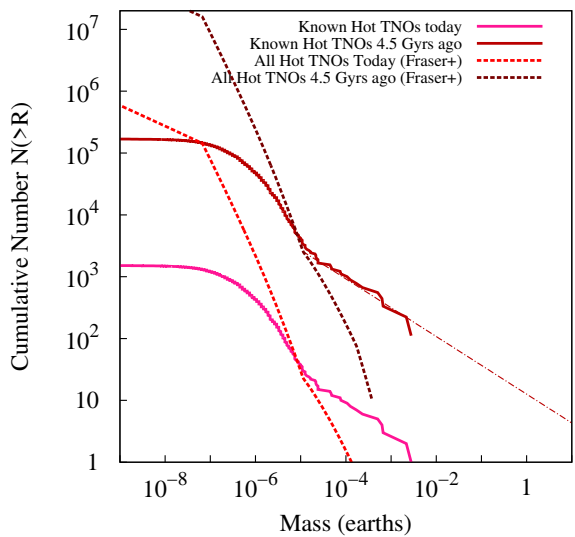}
  \caption{The present day mass-number Hot Trans-Neptunian Object population, and the inferred primordial population.  Extrapolating the high mass end of the mass-number relation to larger sizes implies the original population may have included $\sim 10$~Earth-mass protoplanets.  
  }
  \label{fig:scatpop}
\end{figure}

\section{Splitting of Cold Classical Binaries}

\label{sec:BinarySplits}
The Cold Classical Kuiper belt contains a large number of long period $\left( \gtrsim 1 \rm{yr} \right)$ binaries, which are thought to have formed in situ in the Kuiper belt (e.g., \citealt{peti08,park11}). Due to their low orbital binding energies, these binaries are vulnerable to splitting when large objects fly by.  This vulnerability was first reported by \citet{peti04}. \citet{peti04} conclude the vulnerability was unimportant because they assumed on a small number of large objects, based on the \citet{durd00} theoretical size-number distribution.  At the present, binaries are indeed safe from splitting because of the small number of large objects \citep{2014AJ....147....2S}.  However, today's Hot TNOs are thought to have originated in a disk of a tens of Earth masses interior to Neptune's present day orbit \citep{malh93,levi08}, of which less than a tenth of a percent persists today \citep{2014ApJ...782..100F}. When this material was ejected from the Solar System, most of it passed through the Cold Classical Kuiper belt region. Large embryos scattered through the Kuiper belt risked gravitationally splitting ultra-wide Cold Classical binaries formed in situ.

Here we place limits on the number of large embryos formed interior to Neptune in the early Solar System based on the requirement that most Cold Classical binaries stay bound. We derive the binary preservation requirement from a subset of Cold Classicals with a high observational completeness to binary detection. Overall, $\gtrsim 30\%$ of Cold Classical targets are \emph{detected} binaries \citep{noll08}, but the true fraction is likely much higher. The fraction of bright (H$>$6.1) cold classical targets that are binaries is consistent with 100\% \citep{noll14}. Recently \citet{2017NatAs...1E..88F} identified a subset of Cold Classicals with blue colours, observed them with the Hubble Space Telescope, and found that seven out of eight targets are binary pairs. \citet{2017NatAs...1E..88F} hypothesised that these objects formed at $\sim$ 38 AU, just inside the present day Kuiper belt, and were gently pushed into today's Classical region (42--48 AU) during Neptune's migration. Here we assume that these blue binaries were subject to similar binary disruption rates as the overall Cold Classical population. This assumption is not dependent on the timing of the push out because large scattered embryos would have passed through the blue Cold Classicals' putative formation region as well as the Classical region. Red Cold Classicals may have similar binary fraction rates, but our constraint holds whether or not they do.

\subsection{Analytical estimate of binary splitting timescale}
First we estimate the binary splitting timescale. When a large embryo, scattered into the Kuiper belt, flys by a cold classical binary, the tidal force on the binary is 
\begin{equation}
F_t = \frac{G\memb2M_ba_b}{b^3},
\end{equation}
where $b$ is the impact parameter, $a_b$ is the semimajor axis of the binary, and $\memb$~and $M_b$~are the masses of the large scattered embryo and each binary component (approximated as equal mass) respectively. For an encounter velocity $v_e$, the encounter time is $t_e \sim 2 b/v_e$, and the velocity change between the binary components is 
\begin{equation}
\delta v \sim \frac{4 G \memb a_b}{b^2 v_e}.
\end{equation}

A binary will unbind if the velocity change exceeds the orbital velocity ($\delta v > v_b = \sqrt{G2M_b/a_b}$), which translates to a critical impact parameter,  
\begin{equation}
\label{eqn:b}
\frac{b_{\rm split}}{a} < 2 \sqrt{\frac{na}{v_e}\frac{\memb}{M_\odot} \frac{n}{n_b}}.
\end{equation} 
where $n_b$ is the orbital frequency around the binary, $a$ is the semi-major axis around the Sun, $n$ is the orbital frequency about the Sun, and $M_\odot$ is a solar mass. For example, an Earth mass embryo approaching at the Keplerian velocity will unbind the binary if comes within 0.1\% of the binary's semi-major axis about the Sun, or about 0.04 AU. Note that the encounter velocity $v_e$ will cancel in our final expression. In terms the binary semi-major axis $a_b$, 
\begin{equation}
\frac{b_{\rm split}}{a_b} < \sqrt{2} \sqrt{\frac{n_ba_b}{v_e}\frac{\memb}{M_b}}.
\end{equation}

Using a particle-in-a-box formalism (e.g., \citealt{gold04}), we estimate the splitting timescale as the timescale between encounters of the binary and the large embryo,
\begin{equation}
\label{eqn:tsplit}
\tau_{\rm split} = \frac{1}{f} \frac{2 \pi a \Delta a 2h}{\pi b_{\rm split}^2 v_e}
\end{equation}
\noindent where $\Delta a$ is the width of the Cold Classical belt, $h$~is the height of the Cold Classical belt, and  $f$~is the fraction of the time objects scattered into the Kuiper belt spend within the Cold Classical region. Substituting Eqn. \ref{eqn:b} into Eqn. \ref{eqn:tsplit} yields
\begin{align}
\label{eqn:tsplit2}
\tau_{split} &= \frac{1}{f} \frac{h}{a} \frac{\Delta a}{a} \frac{M_\odot}{\memb} \frac{n_b}{n} n^{-1}  \nonumber \nonumber\\&=\frac{20 \rm Myr}{f} \left(\frac{h/a}{0.06}\right)\left( \frac{\Delta a/a}{0.13}\right)\left( \frac{M_\odot/\memb}{3e6}\right)\nonumber\\&\times\left( \frac{n_b/n}{17.5} \right)\left(\frac{0.02 yr^{-1}}{n}\right).
\end{align}

\subsection{Estimating exposure of Cold Classicals to large scattered embryos}
\label{subsec:f}

The fraction $f$ of time objects scattered into the Kuiper belt spend within the Cold Classical region in Eqn. \ref{eqn:tsplit2} depends on the orbits of the embryos through the Kuiper belt and their overlap with the orbits of Cold Classicals. We estimate $f$ using the $N$-body simulations described in section \ref{sec:pastnpresent}. 

In the three migration simulations, active particles -- those that have not been ejected or collided with a planet or the Sun -- spend an average of $f = \left( 0.8\%, 1.1\%, 0.8\%\right) = 0.9\%$~of their time within the Cold Classical volume.  This quanitity was computed by dividing the total time spent by all particles in the Cold Classical volume by the total lifetime of all particles.  We also find that, averaged over 4 Gyr, at any given time $f=$0.9\% of the Hot TNOs that remain bound to the Sun are passing through the Cold Classical region.  In the static case, we find a similar $f = 0.8\%$.

Substituting $f=0.009$ into Eqn. \ref{eqn:tsplit2} yields an unbinding time for binaries of 
\begin{align}
\label{eqn:tsplit3}
\tau_{\rm split} = & 2.2~{\rm Gyr} \left(\frac{.009}{f}\right)\left( \frac{h/a}{0.06} \right)\left(\frac{\Delta a/a}{0.13}\right)\left( \frac{M_\odot/\memb}{3e6}\right)\nonumber\\
&\times \left( \frac{n_b/n}{17.5} \right)\left(\frac{0.02 yr^{-1}}{n}\right).
\end{align}
, where $M_\odot/\memb$ 3e6 corresponds to a Mars mass perturber.

\subsection{Numerical simulations of binary splitting timescale}

We test the validity of the ultra-wide binary splitting time equation (Eqn. \ref{eqn:tsplit3}) using another set of $N$-body simulations. We conduct 4 Gyr simulations beginning with Neptune on its current orbit, ten binary KBOs, and ten large embryos. Each binary has an initial orbit around the Sun drawn uniformly in the intervals $a =$ 42--48 au, $e =$ 0--0.05, and $i =$ 0--4.5$\degree$ and equal mass components each with $m_b = 10^{21} \rm{g}$ $\left( \sim 10^{-7} m_\oplus \right)$, on $a_b = 10^{10} \rm{cm} \left(\sim 10^{-3} \rm{au}\right), e = 0, i = 0$~binary orbits, comparable to the mutual orbits of the ultra-wide binaries \citep{park11,2017NatAs...1E..88F}.  
Each perturbing embryo has mass of $\memb$, $q = 30$, $a$ drawn randomly and uniformly from 40--140 au, and $i$ drawn randomly and uniformly 0--4.5 $\degree$.

\begin{figure}
\includegraphics[width=0.33\textwidth, angle=270, trim = 0 0 0 0, clip]{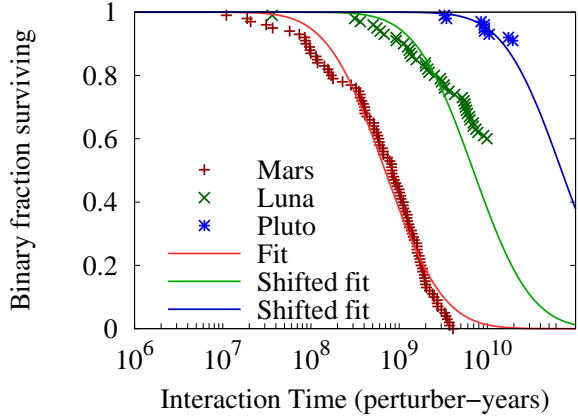}
  \caption{Binaries are susceptible to disruption by large embryos scattered into the Kuiper belt. Fraction of ultra-wide binaries that remain bound as a function of the total time spent by perturbers of Mars mass (red line), Luna mass (green line), and Pluto mass (blue line) as Hot TNOs. The red line is fitted to the red points (see text for details) and shifted by $M_{\rm Mars}/\memb$ for Luna and Pluto mass. The Interaction Time is the sum of the lifetime of the perturbers and is computed after each binary disruption.}
  \label{fig:splittime}
\end{figure}

For each embryo mass ($\memb$ = Pluto, Luna, Mars), we conduct 10 simulations. We do not consider Earth mass embryos here, because they cause secular excitation of the belt (\textsection \ref{sec:coldexcite}) on a timescale comparable to the binary splitting time. The secular excitation can lead to splitting of the binary by Neptune instead of the embryo, making the simulation unsuitable for testing Eqn. \ref{eqn:tsplit3}. Binaries are exposed to embryos until the embryos are ejected or, more rarely, collide with Neptune. We define the Interaction Time as the sum of the lifetime of the embryos at the time of the binary's disruption. For example, if a binary is disrupted at 1 Gyr when nine embryos remain and the other embryo was ejected at 0.5 Gyr, the Interaction Time would be 9.5 Gyr. In Fig. \ref{fig:splittime}, we plot the fraction of remaining binaries as a function of Interaction Time. If the total Interaction Time exceeds the binary splitting time (Eqn. \ref{eqn:tsplit3}), the binary will split.

We find that our estimated binary splitting timescale (Eqn. \ref{eqn:tsplit3}) is a reasonable match to the simulated mean survival time of an ultra-wide binary when massive embryos are present in the scattered disk (Fig. \ref{fig:splittime}).  The best fit log-normal distribution to simulated Interaction Times fit to the Mars-mass case has a centre of $10^{8.84 \pm 0.01} = 0.7$ Gyr with a width of $\sigma = 0.51 \pm 0.02$. The centre agrees to within an order of magnitude with the estimated value of 2.2 Gyr from Eqn \ref{eqn:tsplit3}. We shift and overplot the Mars best-fit line by $\memb^{-1}$ to compare to the Luna and Pluto simulations and find the analytically estimated mass scaling is qualitatively consistent with the simulations.

\subsection{Limits on perturber masses}
Next we combine Eqn. \ref{eqn:tsplit3} with observed estimates of the binary survival rate \citep{2017NatAs...1E..88F} to place limits on the number of massive embryos in the inner disk of the early Solar System. To place these limits, we need also need to know how long the Hot TNOs remain in the Kuiper belt, which sets the Interaction Time.  We use the survival times from the simulations in section \ref{sec:pastnpresent}, plotted in figure \ref{fig:scatlifetime}. 

\citet{2017NatAs...1E..88F} found that seven targets in their sample were binaries and one was a single. We do not know the fraction of Cold Classicals that formed as binaries, but the observed binary fraction is a lower limit on their survival rate. Assuming an initial fraction of $100\%$ allows the most generous allowance for splitting.  We consider embryo masses of $6 \times 10^{24}$, $6 \times 10^{25}$, $6 \times 10^{26}$, and $6 \times 10^{27}$~g, which we refer to as Plutos, Lunas, Marses, and Earths respectively.  For each embryo mass, we consider $N=1, 2, 3...$ embryos of that mass and perform 10000 Monte Carlo simulated surveys.  For each of 10000 simulations, we calculate the total Interaction Time by drawing a survival time for each of $N$ embryos from 3000 particle lifetimes recorded in the simulations of Neptune's outward migration (Fig. \ref{fig:scatlifetime}). 
Because only $50\%$~of objects from the primordial disk pass through the Kuiper belt (Section \ref{subsec:f}), we randomly set each lifetime to 0 with a $50\%$~probability.  

\begin{figure}
\includegraphics[width=0.33\textwidth, angle=270, trim = 0 0 0 0, clip]{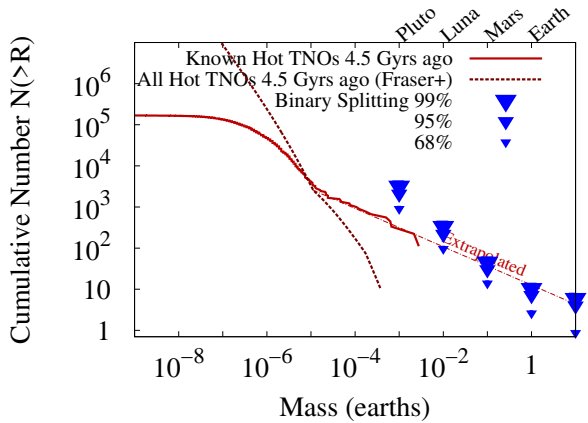}
  \caption{Survival of ultra-wide binary cold classical Kuiper belt objects imposes significant limits (triangles) on the masses of protoplanets and planetesimals formed interior to Neptune and scattered into the Hot TNO population. Overplotted are the inferred initial Hot TNO population, made by scaling the known present-day population (solid red line) and the present-day population inferred by \citet{2014ApJ...782..100F} (dotted red line) by the inverse of the depletion factor inferred in  \textsection \ref{sec:pastnpresent}. Scaling up the present-day large bodies to the inferred primordial population is cosistent with the limit derived here, but extrapolation of the mass-number distribution to larger sizes would not be.}
  \label{fig:comptoscatdisk}
\end{figure}

Three of the binaries (2001 QW$\textsubscript{322}$, 2001 XR$\textsubscript{254}$, and 2003 UN$\textsubscript{284}$) have known masses and semimajor axes \citep{2009Icar..200..627G,park11}.  Four (2002 VD$\textsubscript{131}$, 2003 HG$\textsubscript{57}$, 2014 UD$\textsubscript{225}$, 2016 BP$\textsubscript{81}$) have reported absolute magnitudes \citep{2018ApJS..236...18B} and projected separations \citep{2017NatAs...1E..88F}.  To convert the absolute magnitudes into masses, we assume the bodies are spheres with an albedo of 0.16 \citep{2014ApJ...782..100F}, a density of $0.6$~g/cm$^3$~\citep{2013ApJ...778L..34B}, and the secondary to primary brightness ratio from \citet{2017NatAs...1E..88F}.  To convert projected separations into semimajor axes, we assume the binaries have eccentricities of 0.5 (the average of the binaries in \citealt{park11}, and of the ones in this sample with known eccentricity).  We assume the binaries are roughly flat, so the average actual separation is $\sqrt{2}\times$~the projected separation.  A binary with eccentricity $0.5$~has a median separation of $1.22 a$, so we convert projected separations into semimajor axes by multiplying by $\sqrt{2}/1.22$.  For 1999 OE$\textsubscript{4}$, the single object for which estimating the binary parameters would be extremely fraught with difficulty, we randomly assign it the properties of one of the other seven binaries, with probability proportional to that binary's orbital period and hence inversely proportional to its splitting time (eqn. \ref{eqn:tsplit3}).  Because a split binary is twice as likely to be discovered by an observational survey, we only include a 1999 OE$\textsubscript{4}$~analogue in 50\% of the Monte Carlo trials.  For each binary, we draw a splitting time. If it is less than the Interaction Time, the binary splits.  Those trials with 7 or 8 binaries among the detections are counted as compatible with the Solar system.  We then determine the $68\%$, $95\%$, and $99\%$~upper limits on the number of perturbers where $68\%$, $95\%$, and $99\%$~of trials are not compatible with the Solar system.  The results are plotted in figure \ref{fig:comptoscatdisk} and summarised in table \ref{tab:bin}.   Although the total number of systems is small, it is somewhat worrying to note that the systems with estimated masses and orbits are more tightly bound than those systems with directly measured masses and orbits.  Therefore we repeat the exercise using the seven ultra-wide binaries with known masses and orbits from \citep{park11}, which produces limits that are $30\% - 50\%$~tighter.

\begin{table}
  \begin{tabular*}{\textwidth}{|l | r r r|}
    \cline{1-4}
      Mass & 68\% & 95\% & 99\% \\
    \cline{1-4}
      Earth & 3 & 9 & 14 \\
      Mars & 16 & 38 & 61 \\
      Luna & 111 & 276 & 452 \\
      Pluto & 1040 & 2624 & 4297 \\
     \cline{1-4}
  \end{tabular*}
  \caption{Statistical limits on the initial number of bodies within the primordial disk, as constrained by the survival of ultra-wide binaries within the Cold Classical Kuiper belt.}
  \label{tab:bin}
\end{table}

We infer the primordial population by scaling the present day population up by the inverse of the fraction of objects in our simulations in \textsection \ref{sec:pastnpresent} that survived for 4 Gyrs. We plot the resulting distributing (orange dotted) in Fig.  \ref{fig:comptoscatdisk}. Our upper limits on the number of Pluto mass objects is greater than number of Plutos expected from the inferred primordial population.  However, if we extrapolate the inferred primordial size distribution to larger masses, it lies above our upper limits.  Therefore our upper limits imply a cut-off or quenching of planetesimal-formation at large sizes. It has previously been suggested \citep[e.g.,][]{2008ssbn.book..335B}
that there may be an upturn in the size distribution at sizes greater than $R \sim 300 \rm{km}$ (i.e., the dotted curve in Fig. \ref{fig:comptoscatdisk} lies below the pink curve, though only Pluto and Eris are formally incompatible with the size distribution measured for the hot disk by \citealt{2014ApJ...782..100F}, given the measurement uncertainties).  If such an upturn does exist, it must have ended at the size of the largest objects (Pluto, Eris) present today. We will further discuss the implications of our limits for planetesimal-formation models in \textsection \ref{sec:discussion}

\section{Cold Classical Kuiper Belt Orbital Excitation}

\label{sec:coldexcite}

In the Cold Classical Kuiper belt, the low inclination, higher eccentricity orbits are unoccupied, despite being stable for the age of the Solar System (e.g., \citealt{daws12}, Fig. 3 therein). The Cold Classical KBOs (CCKBOs)' low eccentricities are evidence of their in-situ origin (e.g., \citealt{baty11}). These low eccentricities also imply that CCKBOs experienced  little or no dynamical excitation from their fellow trans-Neptunian objects. Large embryos -- invoked to explain features of the TNO populations -- have a propensity to excite the CCKBOs \citep{glad06,lyka08,yeh09}. Here we will use the CCKBOs' low eccentricities as a limit on primordial population of large embryos formed interior to Neptune and scattered through the Kuiper belt.  

\begin{figure}
\includegraphics[width=0.33\textwidth, angle=270, trim = 0 0 0 0, clip]{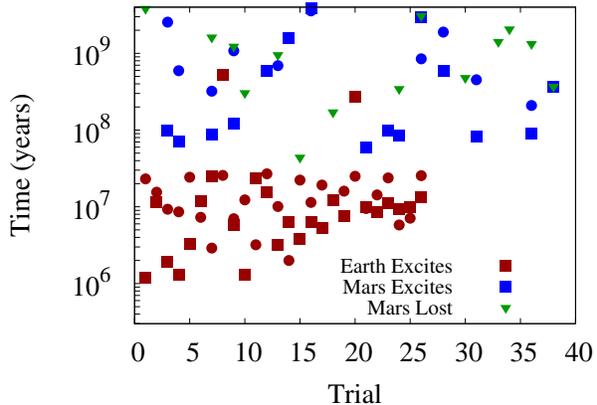}
  \caption{Earths and Mars can excite the CCKBOs on timescales shorter than the age of the Solar System. Distribution of eccentricity excitation times (red squares) and inclination times (red circles) for the Cold Classical Kuiper belt by a single Earth mass perturber, eccentricity excitation times (blue squares) and inclination excitation times (blue circles) for a single Mars mass perturber, and the survival times of the Mars mass perturbers (green triangles).}
  \label{fig:inxcite2}
\end{figure}

To investigate the effect of large embryos on CCKBO orbits, we perform simulations with Neptune on its current orbit; 100 test particles representing CCKBOs from 42.5 to 46 au with initial $e = 0.001$~and $i = 0.05 \degree$; and a single perturber with initial periapse $q = 30~\rm{au}$, semimajor axis $a > 30~\rm{au}$~chosen randomly with probability $\propto a^{-2.5}$, and $i$~chosen randomly and uniformly between $0 \degree$~and $40 \degree$, roughly the range of the hot populations \citep{brow01}.  
We perform 39 simulations with Mars mass embryos and 26 with Earth mass embryos. 
We do not investigate Luna mass and smaller perturbers because, based on the results from Mars, we expect the limits on excitation time to be computationally unfeasible. 
We define the excitation time $t_{\rm{excite}}$~to be the first time when $ \geq 10 \%$~of the CCKBOs have $i \geq 7 \degree$ and/or $e \geq 0.1$. We record the time at which this excitation happens for eventual comparison to the amount of time the embryo spends in the Cold Classical region.  The results are plotted in Fig. \ref{fig:inxcite2}.  Earths and Mars can excite the CCKBOs on timescales shorter than the age of the Solar System.  \citet{2014Icar..232...81M} found that it is possible for the late-stage migration of Neptune to remove Cold Classicals at $e \gtrsim 0.05$~interior to the 7:4 mean-motion resonance with Neptune.  Therefore, we also calculate limits considering only the inclination excitation.

\begin{figure}
\includegraphics[width=0.33\textwidth, angle=270, trim = 0 0 0 0, clip]{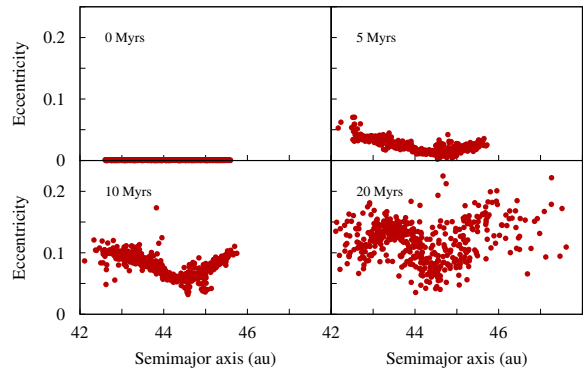}
  \caption{Excitation of a Cold Classical Kuiper belt by an Earth mass perturber, which began on an $a = 63$~au, $e = 0.524$, $i = 13 \degree$~orbit.  Secular interactions coherently drive the Cold Classical Kuiper belt to high eccentricity, though the eccentricities become incoherent as the perturber's orbit is changed by close encounters with Neptune.}
  \label{fig:secular}
\end{figure}

This excitation is driven primarily by secular interactions.  An example, which begins with an Earth mass perturber on an $a = 63 \rm{au}$, $e = 0.524$, $i = 13 \degree$~orbit, is plotted in figure \ref{fig:secular}.  Excitation time is expected to scale as $\tau_{excite} \propto n\mu$~\citep{murr99}.  Exact analytic estimates are difficult as the prefactor can be quite large when the perturber has a semimajor axis similar to the Cold Classicals, and the semimajor axis of the perturber changes after each encounter with Neptune.  Nonetheless, the typical excitation time in the simulations for an Earth is comparable to the $n\mu$~above of $\sim 10^{7}$~years.  For lower mass pertubers, competition with precession induced by Neptune should suppress the excitation.

To place limits on the masses of large embryos that pass through the Cold Classical region, we need to know whether the embryos stay long enough to excite the Cold Classicals. Proper treatment of the ejection of large embryos requires some care. Mars mass embryos often collide with Neptune or are ejected before they excite the CCKBOs because they take longer to do so (Fig. \ref{fig:inxcite2}). Our Neptune-only simulations do not give realistic lifetimes because typically Jupiter is the planet that ejects planetesimals, which are handed off from Neptune. Including all four giant planets in our simulations featuring massive embryos is complicated by the fact that Uranus and Neptune lie just inside the 2:1 mean motion resonance\footnote{Oddly, Uranus and Neptune lie within the deficit of period ratios slightly less than 2 observed by Kepler \citep{liss11,fabr14}.}. Interactions between the large embryo and Neptune or Uranus can cause Uranus and Neptune to cross the 2:1 resonance, increase Neptune's eccentricity, and cause the $\nu_8$~or $\nu_{18}$~ secular resonances to enter the Cold Classical Belt. This event can disturb the CCKBOs but does not serve as a robust constraint because it might be avoided by starting Neptune and/or Uranus on somewhat different orbits. As a simple treatment, we use Neptune-only simulations to measure CCKBO excitation and draw embryo lifetimes from our simulations in Section \ref{sec:pastnpresent}.

\begin{figure}
\includegraphics[width=0.33\textwidth, angle=270, trim = 0 0 0 0, clip]{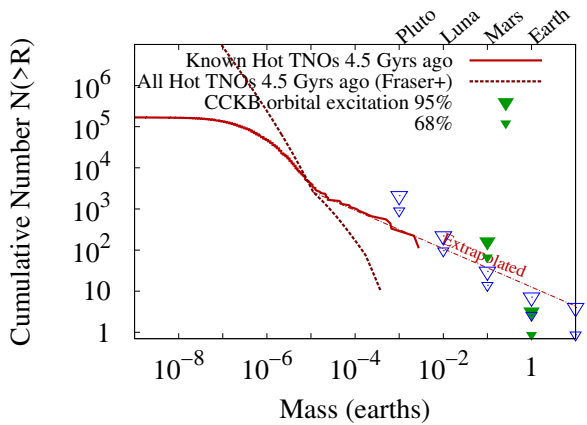}
  \caption{Limits on the primordial population of trans-Neptunian bodies from dynamical non-excitation of the Cold Classical Kuiper belt (solid green triangles) and  ultra-wide binary survival (hollow blue triangles), compared to the inferred initial Hot TNO population, made by scaling the known present-day population (solid red line), and the present-day population inferred by \citet{2014ApJ...782..100F} (dotted red line) up by the inverse of the fraction of particles from our simulations in \textsection \ref{sec:pastnpresent} that survived for 4 Gyrs). For Earth mass perturbers and larger, the non-excitation of the Cold Classical Kuiper belt is a stronger constraint than the survival of the ultra-wide binaries (\textsection \ref{sec:BinarySplits}).}
  \label{fig:excitelimits}
\end{figure}

As in \textsection \ref{sec:BinarySplits}, we assume only $50\%$~of the primordial population formed interior to Neptune is scattered in the Hot TNO population, setting half of embryos' lifetimes to zero to reflect this.  For each embryo, we draw a lifetime from the four planet simulations from Section \ref{sec:pastnpresent}, and compare to the excitation time from a randomly selected Neptune-only excitation time simulation.  If the embryo in the Neptune-only simulation was lost in less time than the drawn lifetime, we redraw a Neptune-only simulation until this is no longer true.
If the Cold Classicals are excited in less time than the lifetime of the embryo, that trial is incompatible with the CCKOs.  We repeat this procedure for $N$ embryos, and if any of them excite the belt, that trial is incompatible with the CCKBOs.  We perform 1000 trials to determine the $68\%$~and $95\%$~upper limits on embryo mass  The results are plotted in figure \ref{fig:excitelimits} and summarised in table \ref{tab:excite}.

\begin{table}
  \begin{tabular*}{\textwidth}{|l | r r r r r r|}
    \cline{1-7}
      Mass & 68\% & 95\% & 99\% & 68\% (i) & 95\% (i) & 99\% (i) \\
    \cline{1-7}
      Earth & 1 & 4 & 7 & 2 & 6 & 9\\
      Mars & 77 & 202 & 320 & 528 & 1426 & 2197 \\
     \cline{1-7}
  \end{tabular*}
  \caption{Statistical limits on the initial number of bodies within the primordial disk, as constrained by the non-excitation of the Cold Classical Kuiper belt.  The limits with (i) consider only inclination excitation.}
  \label{tab:excite}
\end{table}

For Earth mass perturbers and larger  in the progenitor population of the Scattered disk, the non-excitation of the Cold Classical Kuiper belt is a stronger constraint than the survival of the ultra-wide binaries (\textsection \ref{sec:BinarySplits}).  The mass scaling for exciting the CCKBOs is strong, and thus CCKBOs' low eccentricities provide a weaker constraints on Mars mass perturbers than binary splitting.  Although they considered a somewhat different scenario, it is interesting to note that this limit is roughly in line with the result of \citet{baty12}, who found that interactions with a third ice giant, which was ultimately ejected, excited the Cold Classicals in more than $95\%$ of their trials (i.e., $N \lesssim 1$ for a Neptune-mass perturber).

\section{Survival of Resonant Objects}

\label{sec:reslives}

Roughly $15\%$~of Hot TNOs are in mean-motion resonances with Neptune \citep{glad12}, implying a total mass of $\sim 10^{-3} M_{\oplus}$.  The mass of objects occupying resonances today correspond to roughly $0.01\%$~of the total primordial population \citep[e.g.,][]{nesv16}
, and hence the resonances must be able to retain at least that fraction of the primordial population over the Solar system's history (i.e., assuming 100\% of objects were initially in resonance gives us the maximum allowed depletion).
With the current configuration of giant planets and without additional perturbations, about $20\%$~of resonant objects survive for 4 Gyrs, with the rest lost due to resonant diffusion (e.g., \citealt{tisc09}). 
\citet{glad12} found that including Pluto (i.e., including it as a body with mass instead of a test particle) in their simulations
lead to the loss an additional few percent of the initial resonant population, suggesting that the survival of the resonant objects that remain today can provide a limit on the primordial population of large bodies. Here we investigate whether massive embryos scattered into the Hot TNO population would completely deplete the resonances.

Eccentricity-type resonances with Neptune are characterised by a resonance angle
\begin{equation}
    \phi = j \lambda - \left(j-k\right) \lambda_N - k\varpi,
\end{equation}
where $\lambda$ is the mean longitude of the KBO, $\lambda_N$ is the mean longitude of Neptune, $\varpi$ is the longitude of periapse of the KBO, and $j$ and $k$ are integers. 
Over time, $\phi$~oscillates around a centre of libration $\phi_0$~to a maximum amplitude of $\phi_0 + \Delta \phi$~and a minimum of $\phi_0 - \Delta \phi$.  A perturber flying by the resonant object will impulsively change $\lambda$~and $\varpi$, and consequently $\Delta \phi$.  A perturber flying by Neptune will impulsively change $\lambda_N$~and hence $\Delta \phi$.  If the fly by results in $\Delta \phi \gtrsim \pi$, the resonant angle will circulate rather than librate, and the object will be lost from resonance. Moreover, TNOs can be dislodged from resonance when a perturber flys by Neptune. Because perturbers scattered into Hot TNO population are put onto their orbits by encounters with Neptune and because of Neptune's strong gravitational focusing, we focus here on effect of encounters between Neptune and the perturber on resonant KBOs. We focus on the 3:2 resonance because it has the best observationally estimated total mass \citep{2016AJ....152...23V}.

\begin{figure}
\includegraphics[width=0.33\textwidth, angle=270, trim = 0 0 0 0, clip]{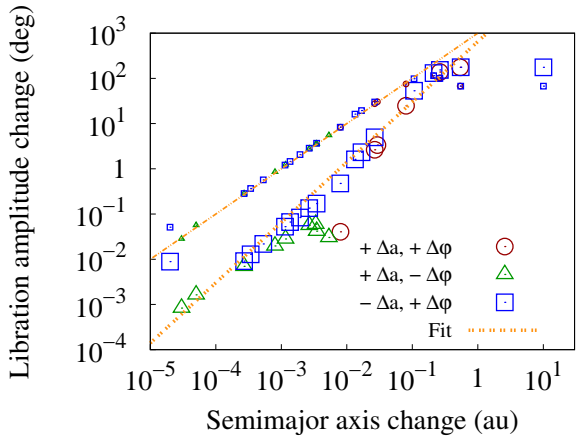}
  \caption{The mean change in libration amplitude (thick lines, large points), and standard deviation thereof (this lines, small points), among 506 synthetic 3:2 resonant objects, after a semimajor axis change $\left( \Delta a_N\right)$ is applied to Neptune.  The fit is to the mean change equation \ref{eq:deltaphi} (thick orange line), while the fit to the spread is equation \ref{eq:deltadeltaphi} (thin orange line).  
  When a positive $\Delta a$~produces a positive libration amplitude change $\Delta \phi_0$, the points are plotted in red,when a positive $\Delta a_N$~produces a negative $\Delta \phi_0$, the  points are green, and when a negative $\Delta a_N$~produces a positive $\Delta \phi_0$, the points are plotted in blue.  
  }
  \label{fig:dphivda}
\end{figure}

To assess how the libration amplitude $\left(\Delta \phi_0\right)$~changes due to an instantaneous change in Neptune's semimajor axis $\left(\Delta a_N\right)$, we perform $N$-body simulations of 506 test particles in 3:2 resonance with Neptune\footnote{These Plutinos were generated by placing test particles randomly and uniformly from $a = 40 \rm{au}$~to $a = 41 \rm{au}$, $e = 0.0$~to $e = 0.1$~and $i = 0 \degree$~to $i = 15\degree$, with $\omega, \Omega, M$~from $0 \degree$~to $360 \degree$.  Particles were then evolved with Poynting-Robertson drag with $\beta = 0.01$~for $10^8$~years, and particles librating with $\Delta \phi_0 < 165 \degree$~after $10^8$~years were retained}. 
In each simulation, we instantaneously displace Neptune. We plot the mean and dispersion of the change in libration amplitudes in Fig. \ref{fig:dphivda}. 

The dispersion among the 506 test particles for a kick of $\Delta a_N$~is well fit by
\begin{equation}
 \delta \Delta \phi_0 = 1000 \degree \left(\frac{\Delta a_N}{1 \rm{au}}\right).
 \label{eq:deltadeltaphi}
\end{equation}

The mean $\Delta \phi_0$~is well fit by 
\begin{equation}
 \left| \Delta \phi_0 \right| = 650 \degree \left|\frac{\Delta a_N}{1 \rm{au}}\right|^{4/3}.
 \label{eq:deltaphi}
\end{equation}

Neptune experiences these changes in semimajor axis during close flybys of perturbers in the Hot Trans-Neptunian population.  Each orbit, the perturber flies by Neptune at closest approach $b$, resulting in a force $F = GM_nm_p$~over a time $t = b/v_e$, where $v_e$~is the encounter velocity.  This impulse results in a velocity change
\begin{equation}
 \delta v = \frac{F \times t}{M_n} = \frac{Gm_p}{b^2}\frac{b}{v_e}.
  \label{eq:widthdisp}
\end{equation}

The corresponding energy change is
\begin{equation}
 \delta E = 2 M_N v_e \delta v,
 \label{eq:widthmean}
\end{equation}

and a corresponding semimajor axis change is 
\begin{equation}
 \Delta a_N = \frac{2 a_N^2}{G M_* M_N \delta E} = 4 \frac{a_N^2}{b} \frac{m_p}{M_*}.
 \label{eq:deltaaenc}
\end{equation}

\begin{figure}
\includegraphics[width=0.33\textwidth, angle=270, trim = 0 0 0 0, clip]{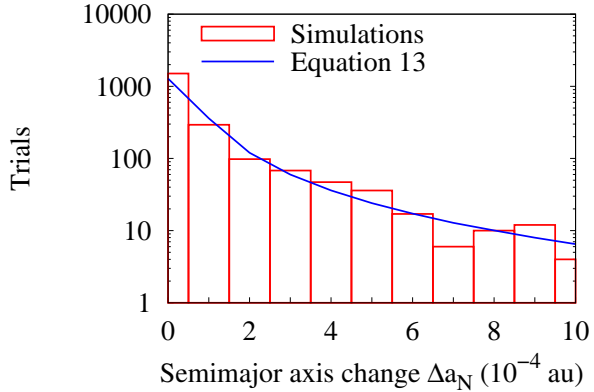}
  \caption{Comparison of the distribution of $\delta a_N$~from equation \ref{eq:deltaaenc}~with uniform encounter distances to an~$N$-body simulation of encounters between Neptune and a Mars mass perturber.}
  \label{fig:jumpfits}
\end{figure}

To verify Eqn. \ref{eq:deltaaenc}, we perform 10 Myr $N$-body simulations of a Mars mass perturber, beginning with $a = 63~\rm{au}$~and $q = 30 \rm{au}$, scattering off Neptune. We measure the change in Neptune's semimajor axis after each encounter.  We transform Eqn. \ref{eq:deltaaenc}~into a distribution of $\Delta a_N$~expected for many encounters by assuming that once each orbit of the perturber there is an encounter with $b$~chosen randomly and uniformly between $0$~and $2 a_N$.
The distribution of those changes is plotted in Fig.  \ref{fig:jumpfits}.  Eqn. \ref{eq:deltaaenc} matches the simulation results, and therefore we can use Eqn. \ref{eq:deltaaenc} below to generate changes in Neptune's semimajor axis due to encounters with the perturber.

\begin{figure}
\includegraphics[width=0.33\textwidth, angle=270, trim = 0 0 0 0, clip]{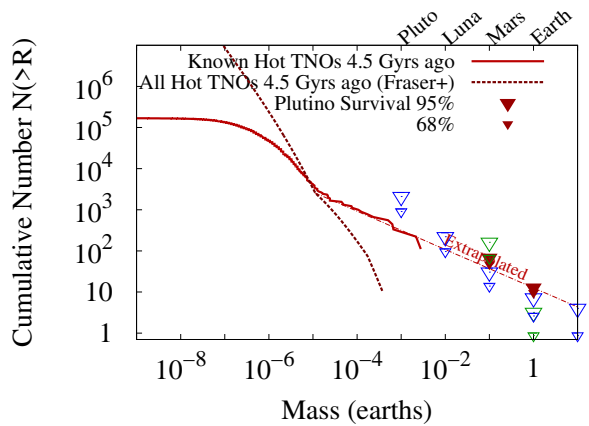}
  \caption{Limits on the primordial population of trans-Neptunian bodies from survival of resonant objects (solid red triangles), non-excitation of the Cold Classical orbits (hollow green triangles) and ultra-wide binary survival (hollow blue triangles), compared to the inferred initial Hot TNO population, made by scaling the known present-day population (solid red line), and the present-day population inferred by \citet{2014ApJ...782..100F} (dotted red line) up by the inverse of the fraction of particles from our simulations in \textsection \ref{sec:pastnpresent} that survived for 4 Gyrs).}
  \label{fig:reslimits}
\end{figure}

As in Sections \textsection \ref{sec:coldexcite} and \ref{sec:BinarySplits}, we determine the upper limit on the primordial number of perturbers ($N$) using Monte Carlo simulations.  For each $N$, we perform 100 draws of lifetimes for each perturber from the simulations described in section \ref{sec:BinarySplits}.  We initialise $10^4$~initial resonant objects with libration amplitude $\Delta \phi_0 = 0\degree$. 
This low initial libration amplitude is likely an underestimate but is a conservative choice. The true distribution of libration amplitudes depends on Neptune's migration history and the resonant KBOs' eccentricities (e.g., \citep{quil06,must11,pike17}), but unless all objects are captured at very large libration amplitude, the reduction in our derived limits will only be of order unity.  Once per orbit of the perturber, we generate an encounter with Neptune and shift Neptune's semimajor axis by an amount given by equation \ref{eq:deltaaenc}, and shift the resonance amplitude of each resonant object by a mean given by equation \ref{eq:widthdisp} and a random component taken from gaussian with a dispersion given by Eqn.  \ref{eq:widthmean}. 
Once the libration amplitude exceeds $180 \degree$, the resonant object is removed from the simulation.  As $\sim 0.01\%$~of the primordial population is in the resonant objects today, we require at least one of the $10^4$ resonant objects to survive for the trial to be successful.  The resulting limits on perturbers are plotted in Fig. \ref{fig:reslimits} and summarised in table \ref{tab:res}.

\begin{table}
  \begin{tabular*}{\textwidth}{|l | r r|}
    \cline{1-3}
      Mass & 68\% & 95\%\\
    \cline{1-3}
      Earth & 11 & 16\\
      Mars & 55 & 85\\
     \cline{1-3}
  \end{tabular*}
  \caption{Statistical limits on the initial number of bodies within the primordial disk, as constrained by the survival of resonant objects.}
  \label{tab:res}
\end{table}

Overall, the survival of resonant objects places less stringent limits on the number of protoplanets that could have formed interior to Neptune than the other constraints we have considered.  One reason that the limits from resonant retention are weaker is that both the survival of blue ultra-wide binaries in \citet{2017NatAs...1E..88F} and the non-excitation of the Cold Classical Kuiper belt require order unity survival rates, while resonant retention requires only $\sim 10^{-4}$~of the objects to survive.  

\section{Comparison to Planet Formation Models}

\label{sec:discussion}

\begin{figure*}
\includegraphics[width=0.67\textwidth, angle=270, trim = 0 0 0 0, clip]{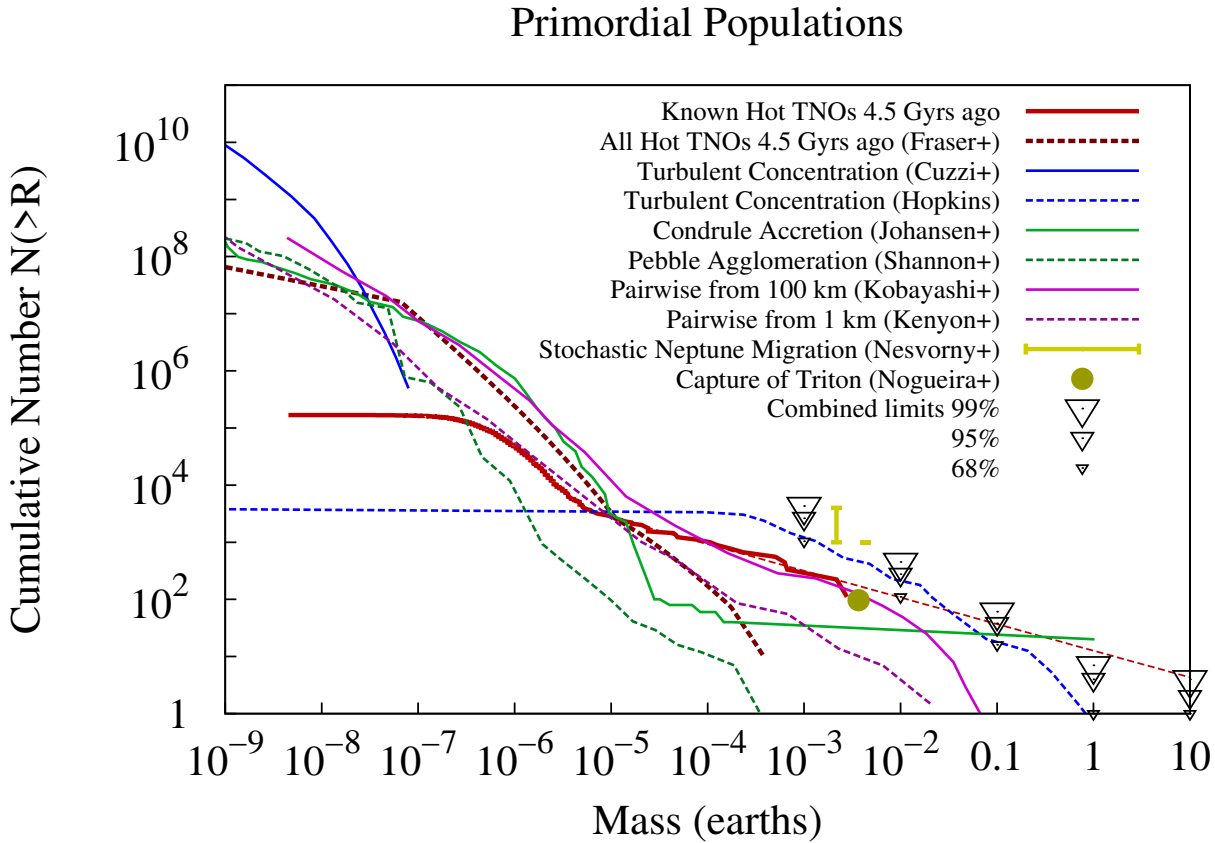}
  \caption{Comparison of the limits derived in this work to theoretical predictions from planetesimal formation and growth models, and models that invoke the dynamical influence of protoplanets in the primordial outer Solar system.  Also included are measurements of the present-day population (red) scaled up by the inverse factor of the depletion seen in dynamical simulations of the scattered disk.  The limits we derive for plutos are compatible with the inferred initial population of plutos.}
  \label{fig:theorycomp}
\end{figure*}

The constraints described in this paper can be compared with the prediction of models of the formation and growth of planetesimals and protoplanets, including for proto-planets more massive than those that survived until today. Here we compare our limits to the predictions of several published models. An important caveat is that many of the models have tunable parameters that affect the resulting planetesimal mass distribution. When a study explores a range of parameters for the model, we take the cases that best matched the TNO formation environment and those models that were presented as the default case (for instance \citet[][]{2016MNRAS.456.2383H} considers planetesimal formation at 0.1, 0.3, 1, 3, 10, 30, and 100 au, so we us 30 au as the closest environment.  They also consider 1 mm, 1 cm, and 10 cm pebbles, so we plot 1 cm pebbles as their default case.) 
The comparisons are summarised in figure \ref{fig:theorycomp}.  We only compare to models that consider planetesimal formation and growth at $20 - 40 \rm{au}$, and ignore studies explicitly considering asteroid formation such as \citet{2009Icar..204..558M,2010Sci...330.1527B}.

To be consistent with the limits we found, a model must either produce only planetesimals less massive than Pluto or a mass-number distribution with a slope no shallower than $dn/dm \propto m^{-a}$, with $a \gtrsim 2$. Traditional models of planetestimal growth in the Kuiper belt region \citep{keny98,orme10,schl11}, which begin with a Minimum Mass Solar Nebula \citep{1977Ap&SS..51..153W} worth of $\sim \rm{km}$~planetesimals produce $a \approx 2$~to a maximum size of $\sim 1000 \rm{km}$ due to the trans-Hill growth timescale \citep{lith14}, making them generally consistent with the limits we find here.  
However, these models are challenging to reconcile with other constraints such as the survival of ultra-wide binaries due to collisional unbinding \citep{park12}, the evolution of exoplanetary debris disks \citep{shan15}, the collisional history of the asteroid belt \citep{bott05}, or the migration of Neptune \citep{gome04}.  
If one instead considers growth starting with $\sim 100 \rm{km}$~bodies \citep[as suggested by][]{bott05,2009Icar..204..558M}~models find a $q \sim 4$~with maximum sizes of $3000 \sim 5000 \rm{km}$~\citep{keny12,koba14}~which is barely consistent with our limits here.

\citet{cuzz08,cuzz10}~developed predictions for the initial mass function of planetesimals produced by the concentration of pebbles in MRI-generated turbulence in a disk.  The general properties of the resulting planetesimals are sensitive to the properties of the gas disk and the properties of the turbulent cascade of power from large scales to small scales, which are not well constrained. As a consequence, it is difficult to compare that mechanism to the mass limits derived here.  Nonetheless, the predicted size-number distributions are generally peaked in a narrow size range at $R = 50 \rm{km} \sim 500 \rm{km}$~which would be generally consistent with the limits here.  An important caveat is that the model only tracks the planetesimals through the  turbulent concentration and contraction stage, and subsequent evolution might modify the size-number distribution.  

\citet{hopk16}~also considered the creation of planetesimals by turbulent concentration of pebbles.  They found top heavy ($q < 4$) mass distributions extending up to Mars to Earth masses, putting their model in tension with the limits we have derived here; however this is only at the 68\% -95\% level.  \citet{hopk16} attributed the difference to \citet{cuzz10}'s assumption that the grains would be sized to optimally couple to the smallest eddies and questioned whether the pebble concentrations in \citet{cuzz10}'s semi-analytic model exceeded what is physically plausible. If the assumptions in \citet{cuzz10} are unrealistic and if \citet{hopk16}'s modeling and assumptions capture all the important physics, then larger numbers of, and better known orbits for the ultra-wide binaries could constrain whether the turbulent concentration mechanism produces excessively large planetesimals that would have disrupted the TNOs.

Pebble Accretion is a promising mechanism for growing planetesimals and protoplanets \citep{2010A&A...520A..43O,2012A&A...544A..32L}.  It substantially enhances the cross section for accreting small particles that are partially coupled to the gas and have their relative velocities damped by gas drag  \citep[see][for a review]{2017AREPS..45..359J}.  \citet{joha15}~considered the evolution of a population of planetesimals growing by pebble accretion.  They began with a size distribution of large bodies derived from simulations of the Streaming Instability \citep{youd05,joha07}, which produces $q \sim 2.8$~with a maximum mass of $10^{23} \sim 10^{24} \rm{g}$~\citep{2016ApJ...822...55S,simo17}.  \citet{joha15} considered the subsequent evolution of the planetesimals in this environment. They performed simulations of a narrow ring at 25 au and found that a single body underwent significant growth, reaching roughly an Earth mass, amounting to roughly one third of the mass in that ring.  If this result for a single $3 M_\oplus$ ring can be extended to a $30 M_\oplus$~disk spanning 20--30 au, the disk would produce $\sim 10$~Earth mass bodies, inconsistent with the limits derived here.  Whether this result is generally applicable to Pebble Accretion depends on the sensitivity of final mass-number distribution to factors such as the initial mass-number distribution, which is yet to be broadly explored.

Similar to Pebble Accretion, the 
collisional pebble agglomeration model of \citet{shan16} also produces a top heavy $q \sim 2$~size number distribution, but with the largest bodies in the $10$s to $100$s~of kilometers in size, dependent on the composition of the pebbles.  Here, we use the code and general setup from that work, but consider a belt from $20-30$~au that begins with $30 M_\oplus$.  With little mass in bodies larger than $\sim 100$~km, it is compatible with the limits we have found.

In addition to testing planetesimal formation models, our limits offer constraints on the dynamical history of the solar system. Several other works have invoked large bodies within the pre-Scattered disk to explain features of the Solar system.  As described in Section 1, \citet{nesv16} invoked 1000-4000 bodies of mass $1.3 \times 10^{25}~\rm{g}$~or 1000 objects of mass $2.6 \times 10^{25}~\rm{g}$~in the pre-Scattered disk to produce the inferred capture rate of resonant objects.  These choices would be incompatible with the survival of the ultra-wide binaries at the 92\%-99.97\% confidence level.

Uranus has an obliquity of $98\degree$, which is not naively anticipated in planet formation models.  One mechanism for producing the tilt is a collision with multiple roughly Earth mass impactors \citep{safr65,morb12}.  A similar size impactor can produce the obliquity of Neptune \citep{lee07,pari11}.  Assuming these impactors came from the primordial Trans-Neptunian population, a single Earth-mass object would be compatible with the limits.  However, an impact with Uranus is quite an unlikely outcome for any given object, which is far more likely to be ejected in the Uranus/Neptune region \citep{levi01,wyat17}.  In the $N$-body simulations of Neptune's outward migration discussed in section \ref{sec:pastnpresent}, $0.6\%$~of the scattered disk impacted Neptune, and $0.2\%$~impacted Uranus.  In the static Solar system history simulations of \citet{shan15a}, $0.3\%$~of the planetesimals hit Neptune, and $0.2\%$~of the planetesimals hit Uranus.  Clearly at late times, this would imply an unacceptably large number of such bodies.  At early times, additional damping provided by the gas disk can substantially increase the fraction of bodies that are accreted by the ice giants \citep{2015A&A...582A..99I}.  However, for giant impacts to generate the obliquity, the size distribution of bodies accreted must have most of the mass in the largest bodies (i.e., $q < 4$) \citep{harr82}, which would be inconsistent with our limits for bodies larger than Pluto.

The origin of Pluto's large moon Charon is often ascribed to a giant impact \citep{mcki84,mcki89,canu11,mcki17}  In the solar system today, the mean time for a collision is $\sim 10^{13} \rm{yrs}$ \citep{canu05}.  This requirement has lead to the inference that the primordial population must have contained a large number of objects with $R \gtrsim 10^3 \rm{km}$ \citep{ster91}.  
However, fewer big bodies would suffice if the velocity dispersion were low. At zero velocity dispersion the collision rate is enhanced by $\alpha^{-3/2}$, where $\alpha$~is the ratio of the physical size of a body to the size of its Hill sphere. A few Plutos with $e=0$ could generate such a collision, and the typical eccentricities at that time are not well known.  Thus, a direct comparison would require a more quantitative prediction of the number of Plutos needed to produce a Charon-forming impact, which is not yet available.

Triton is a large $m \sim 2 \times 10^{25} \rm{g}$~satellite of Neptune, whose inclined, retrograde orbit has been interpreted it was captured. Capture by the tides from a single encounter was originally suggested \citep{mcki84}, but it is exceedingly difficult to dissipate enough energy to bind Triton.  \citet{gold89} suggested collision with an earlier moon could provide the original binding, though the chances of a suitable collision requires an extremely large number of Tritons, and a rather unlikely collision with a small moon. The difficulty in capturing Triton through tides or collisions has lead to the most popular suggestion, that Triton was a member of a binary and was split and captured by Neptune, with the other component carrying off the excess energy \citep{agno06}.  Modelling by \citet{nogu11} found that this scenario required a primordial population of $\sim 97$~Triton-mass binaries, which is compatible with our limits in $\sim 75\%$~of cases.

\section{Conclusions}

\label{sec:conclusions}

We presented limits on the number of Pluto mass and larger protoplanets that could have been present in the primordial planetesimal belt interior to Neptune that was later scattered out to produce today's Hot Trans-Neptunian Objects.  We obtained these limits by considering the survival of the ultra-wide binaries of the Cold Classical Kuiper belt, the dynamical non-excitation of the orbits of the Cold Classical Kuiper belt objects, and the survival of resonant objects to the present day.  We found that if the primordial mass-number distribution extended to mass greater than Pluto, it must have turned downwards to be no more top-heavy than roughly equal mass per log size $\left( dn/dm \propto m^{\lesssim -1.9} \right)$~from the $dn / dm \propto m^{- \sim 1.5}$~size distribution observed today between $10^{-5} M_{\oplus}$~to $10^{-3} M_{\oplus}$ (i.e., inconsistent with an extrapolation of today's mass distribution). A mass-number distribution peaked at the largest masses may not be surprising, this is predicted by some modern planetesimal formation and growth models \citep[e.g.,][]{shan16,2016ApJ...822...55S}.

These results can constrain planetesimal growth and planet formation models, but the lack of consistent predictions from models currently hampers these constraints.  For instance, the turbulent concentration model of \citet{cuzz10} calculates an initial size distribution that is compatible with the limits we have derived, while \citet{hopk16} models turbulent concentration and obtains a size distribution that is marginally consistent with the limits we have derived.  Similarly, the collisional pebble agglomeration model of \citet{shan16}~produces a mass-number distribution compatible with our limits, while the condrule accretion model of \citet{joha15} produced an earth mass object in a ring 0.5 au wide.  Expanded to a $\sim 10~\rm{au}$ wide ring, this distribution would contain far too many Earths to be consistent with our limits.  Growth of planetesimal models by pairwise collisions generically produce only small numbers of large bodies \citep{shan15,lith14}~and so do not conflict with our limits, whether the initial planetesimals are small \citep[as in][]{keny12}~or large \citep[as in][]{koba14}.  We also compare with some dynamical histories, and find the $\sim 100$~Triton mass objects needed by \citet{nogu11}~to capture a Triton around Neptune is consistent with out limits, while the $1000 \sim 4000$~Pluto mass objects needed by \citet{nesv16}~to stochastically migrate Neptune to limit the capture of resonant TNOs is marginally consistent with our limits.
We encourage future works on the formation and growth of planetesimals and protoplanets -- and works which invoke large planetesimals to explain features of the solar systems -- to consider the limits derived here when evaluating their models.

\section{Acknowledgements}
We thank Wes Fraser and Ruth Murray-Clay for useful discussions, and the referee David Nesvorn\'{y} for a review that improved the quality of this manuscript.  This work was partially supported by funding from the Center for Exoplanets and Habitable Worlds. The Center for Exoplanets and Habitable Worlds is supported by the Pennsylvania State University, the Eberly College of Science, and the Pennsylvania Space Grant Consortium.

\bibliographystyle{mn2e}
\bibliography{tno}

\end{document}